# Information-Backed Currency (IBC): Designing a Resilient, Transparent, and Information-Centric Monetary Ecosystem

*Lalit Kumar Shukla[1*]*

*1. Faculty of Physical Sciences, Shri Ramswaroop Memorial University Uttar Pradesh, India
Email: lalitshukla.phy@srmu.ac.in*

Abstract:

*The accelerating digitization of economic activity has made information the dominant driver of market expectations, coordination, and systemic risk. Yet contemporary monetary systems remain anchored in architectures designed for material scarcity, institutional authority, or cryptographic constraint, leaving them increasingly misaligned with information-driven economies. This conceptual paper proposes Information-Backed Currency (IBC) as a novel monetary framework in which verified, high-integrity information functions as the primary source of value creation and monetary stability.*

*Drawing on insights from econophysics, information theory, and cognitive economics, the paper advances the central proposition that economic value emerges when information measurably reduces uncertainty within complex systems. Building on this premise, the study develops an architectural model in which currency issuance is linked to quantified entropy reduction achieved through multi-path information verification, reproducibility assessment, and contextual validation. An ethical governance layer—termed the Dharma Protocol—is introduced to ensure that only socially stabilizing, non-manipulative information qualifies as currency-backing input.*

*The proposed IBC architecture comprises four interdependent layers: information ingestion, verification and validation, ethical oversight, and monetization through a Verification Value Unit tied to uncertainty reduction. By structurally aligning money creation with epistemic integrity rather than scarcity or authority, IBC aims to enhance transparency, reduce volatility arising from information asymmetry, and strengthen trust in digital economic systems.*

*While the framework is intentionally conceptual and non-empirical, it provides a coherent blueprint for re-imagining monetary governance in an era characterized by information abundance, cognitive constraints, and systemic fragility. The paper concludes by outlining policy implications and future pathways for integrating information-centric value mechanisms into existing monetary and regulatory structures.*

## 1. Introduction: The Crisis of Value in Fiat Economies

The 21st-century economy operates on a paradox. While global markets are increasingly driven by the production, circulation, and manipulation of information, the monetary systems that underpin these markets continue to rely on architectures developed for an industrial world defined by material scarcity and centralized authority. Fiat currencies—anchored in state credibility rather than intrinsic value—have proven remarkably adaptable in the past century but now face structural pressures that challenge their long-term resilience.

Three converging dynamics intensify this vulnerability. First, **monetary expansion** and unconventional policy interventions have contributed to inflationary cycles, asset bubbles, and widening wealth imbalances, eroding public confidence in the stability of fiat money. Second, **information asymmetry**—magnified by digital platforms, algorithmic trading, and high-frequency market reactions—has created volatility that policymakers struggle to anticipate or regulate [1]. Third, the rise of **cryptocurrencies and tokenized assets**, although innovative, has not produced a credible alternative to fiat systems; rather, these instruments have introduced speculative risks, governance gaps, and technological fragilities that further complicate economic coordination.

At the core of these disruptions is a deeper epistemic issue: the global economy generates more information than it can validate, interpret, or trust. When currency value depends on expectations, and expectations depend on information, **trust in information becomes a monetary problem**. In an era where information is abundant but its integrity is uncertain, fiat money becomes increasingly exposed to manipulation, misinformation, and systemic shocks.

This paper positions **Information-Backed Currency (IBC)** as a conceptual response to this crisis. Instead of anchoring value to physical commodities or institutional authority, IBC proposes that currency can be stabilized by **verifiable, high-integrity information flows** that reduce uncertainty and enhance collective trust. This move reframes value not as a derivative of scarcity, but as a derivative of **epistemic stability**. By integrating insights from econophysics, information theory, cognitive economics, and ethical governance, IBC seeks to build a more transparent, resilient, and information-centric monetary ecosystem.

The sections that follow develop this proposition by defining the theoretical foundations of IBC, outlining its technical architecture, and evaluating its implications for policymakers, financial institutions, and global digital governance.

This paper is conceptual and architectural in scope, aiming to establish a coherent theoretical framework rather than to provide empirical calibration or policy implementation. Its primary contribution lies in formalizing information integrity and uncertainty reduction as explicit monetary inputs, thereby offering a novel lens for rethinking currency issuance and monetary governance in information-driven economic systems.

## 2. What Is Information-Backed Currency (IBC)?

Information-Backed Currency (IBC) is a proposed monetary framework in which **verified, ethically validated information** serves as the foundational driver of currency issuance, valuation, and stability. Unlike traditional money, which derives value from state authority (fiat), material scarcity (commodity backing), or cryptographic consensus (blockchain-based assets), IBC grounds value in the **epistemic integrity** of the information entering an economic ecosystem.

At its core, IBC rests on a single proposition:
**When information reduces uncertainty in a measurable, transparent, and reproducible manner, it generates economic value.**
This value, once authenticated, can serve as the basis for a stable and resilient currency.

## 2.1 Distinguishing IBC from Existing Monetary Paradigms

### A. Not Fiat:

Fiat money relies on institutional credibility. Its value can be influenced by political decisions, monetary expansion, or shifts in public sentiment [8]. IBC, by contrast, distributes trust across a network of information validators and entropy-reduction metrics, creating value through epistemic stability rather than authority.

### B. Not Commodity-Backed:

While commodities like gold offer scarcity-based stability, they lack adaptability in an information-driven world. IBC ties currency to a scalable and renewable resource: **verified knowledge**.

### C. Not Cryptocurrency:

Conventional cryptocurrencies rely on cryptographic scarcity and consensus mechanisms that do not assess the truth or integrity of underlying information. IBC does the opposite: it is designed to quantify and monetize **authentic information**, not artificial scarcity.

### D. Not Data Tokenization:

Tokenized data assigns market value to data ownership or access rights. IBC assigns value not to the data asset, but to the **information validation process**—the transformation of raw data into *trusted knowledge* that reduces uncertainty.

## 2.2 The Conceptual Basis: Information as Economic Substance

Information becomes economically meaningful when it:

1. **Reduces entropy** (uncertainty) in decision-making [8], [12].
2. **Improves prediction accuracy** or situational awareness.
3. **Enables coordinated behavior** in social, market, or institutional contexts.
4. **Has verified provenance and integrity**.
5. **Can be reproduced or independently validated**.

In this sense, IBC treats high-integrity information as a form of **negentropy**—a stabilizing force that enhances order and predictability within complex systems. Drawing from Shannon's information theory, the framework posits that the reduction of uncertainty produces measurable economic effects [5], [12]. Currency backed by this process therefore gains stability from the **quality** of information, not the **quantity** of data.

## 2.3 Defining the Unit of Value in an IBC System

The IBC system is built around the principle that:

**Value is generated when new information enters the system and is authenticated through multi-layered verification processes.**

This verification transforms data into "information value density," which is quantified through metrics such as:

- **Entropy Reduction Index (ERI)**
- **Provenance Score**
- **Reproducibility Coefficient**
- **Contextual Integrity Rating**

Together, these yield a **Verification Value Unit (VVU)**—a measurable representation of how much uncertainty an information packet removes from an economic environment. Currency issuance in an IBC system is directly tied to cumulative VVUs.

## 2.4 Illustrative Formalization: Entropy Reduction and Monetary Issuance

To clarify the analytical logic underlying Information-Backed Currency (IBC), this subsection presents an illustrative formalization of how verified information generates monetary value through uncertainty reduction. The purpose is not empirical calibration, but conceptual precision and internal consistency.

Let an economic subsystem be characterized by a probabilistic state space {pi}, where uncertainty is quantified using Shannon entropy:

$$H = -\sum_i p_i \log p_i$$

Prior to the introduction of new information, the system exhibits an initial entropy $H_0$. Following the ingestion, verification, and contextual validation of an information event, the updated probability distribution yields a lower entropy $H_1$, with $H_1 < H_0$. The **Entropy Reduction Index (ERI)** is then defined as:

$$HRI = H_0 - H_1$$

The ERI represents the measurable reduction of uncertainty attributable to a specific, verified information input. Within the IBC framework, only information that passes multi-path verification, reproducibility checks, and contextual integrity assessments is eligible to generate a positive ERI.

To translate uncertainty reduction into monetary value, IBC introduces the **Verification Value Unit (VVU)** as an abstract unit of economic weight. In its simplest conceptual form, VVUs may be expressed as a function of entropy reduction and validation quality:

$$VVU = f(ERI, R, P, C)$$

where R denotes reproducibility, P provenance integrity, and C contextual alignment. The specific functional form of f(·) is left open by design, allowing institutional, sectoral, or jurisdictional implementations to calibrate weighting schemes according to domain-specific requirements and governance constraints.

Currency issuance in an IBC system is proportional to the aggregate VVUs generated within a given issuance window. As a result, monetary expansion becomes endogenously linked to epistemic productivity: periods characterized by high-quality, uncertainty-reducing information flows support controlled currency growth, while informational degradation or manipulation naturally constrains issuance.

This formalization highlights the central departure of IBC from conventional monetary paradigms. Rather than treating information as an external influence on prices or expectations, IBC embeds information quality directly into the mechanics of money creation. The framework thus reframes monetary stability as an emergent property of verified knowledge production rather than institutional authority or artificial scarcity.

## 2.5 The Transformation Engine: Truth as a Monetary Input

Traditional monetary systems treat truth as a normative principle; IBC treats it as a **technical asset**. The process of validating information becomes the functional equivalent of mining, minting, or collateral assessment. Economic value emerges through:

- Truth discovery
- Source triangulation
- Error correction
- Contextual alignment
- Ethical safeguard filters
- Verification consensus

Through this architecture, truth functions as both a **resource** and an **economic stabilizer**.

## 2.6 Why Information Works as Backing

Information possesses characteristics ideally suited for a next-generation currency system:

- **Abundant yet not infinitely valuable**—only verified information counts.
- **Scalable**—information ecosystems grow without resource depletion.
- **Transparent**—provenance and validation can be traced.
- **Ethical alignment potential**—moral filters can be integrated into validation protocols.
- **Resilient against manipulation**—when supported by multi-layer verification.
- **Decentralizable**—without relying on trustless cryptography.

In an era defined by digital acceleration and systemic fragility, information—when properly validated—becomes the most resilient, renewable, and ethically governable foundation for currency creation. The following section situates this formal structure within the broader theoretical foundations of econophysics, information theory, and cognitive economics.

## 3. Foundations: Econophysics, Information Theory, and Cognitive Economics

Designing a currency system grounded in information requires a conceptual foundation that transcends conventional monetary theory. Information-Backed Currency (IBC) draws upon three domains—**econophysics**, **information theory**, and **cognitive economics**—to articulate how information can function as an economic substance with measurable, reproducible, and stabilizing properties [3], [4], [5] [10], [12], [17]. These domains provide the mathematical, structural, and behavioral scaffolding necessary to justify a currency anchored in validated information flows.

## 3.1 Econophysics: Value as a Function of Uncertainty Reduction

Econophysics views economic systems as complex, dynamic networks governed by probabilistic interactions rather than deterministic equilibrium [10], [18], [20]. Prices, risks, and expectations fluctuate in ways that resemble physical systems governed by turbulence, entropy, and nonlinear coupling [3], [10]. In this framework:

- **Markets tend toward higher entropy** without stabilizing information inputs.
- **Information operates as a negentropic force**, reducing disorder and guiding agents toward coordinated decisions [8].
- **Volatility is directly correlated with information asymmetry**, making uncertainty measurable in physical terms [1], [9], [17].

Within this paradigm, IBC conceptualizes information as a **stabilizing energy source**. When high-integrity information enters an economic system, it acts analogously to a force that reduces state uncertainty. Thus:

**Economic value becomes a function of entropy reduction.**
This aligns directly with models in statistical physics, where reduced entropy corresponds to increased order, structural stability, and predictable dynamics [3], [7].

IBC therefore positions currency issuance not as an arbitrary policy decision, but as the measurable output of systemic stabilization achieved through verified information.

## 3.2 Information Theory: Measuring the Economic Substance of Information

Shannon's information theory provides the mathematical core of IBC. In this framework:

- Information is defined as the **reduction of uncertainty** in probabilistic states [5].
- Entropy serves as the metric of that uncertainty.
- Reproducibility, redundancy, and noise influence how information affects system dynamics [12], [13].

IBC adopts these principles by quantifying information not by its volume, but by its **uncertainty-reducing capacity** [7]. Three pillars from information theory support this:

### A. Shannon Entropy as Value

Information has economic value when it reduces entropy (H). In IBC, the Entropy Reduction Index (ERI) becomes a core parameter for determining the Verification Value Unit (VVU), the quantitative basis of currency issuance.

### B. Signal-to-Noise Ratio (SNR)

Economic systems are saturated with noise—misinformation, speculation, opinion, and algorithmic distortions. IBC therefore emphasizes high-SNR information that improves decision reliability [5] [12].

### C. Redundancy and Reproducibility

Verified information must be reproducible across independent pathways. This ensures robustness, provenance, and resistance to manipulation [13].

Thus, information only becomes "currency-grade" when it transitions from raw data to **validated, entropy-reducing signal**.

## 3.3 Cognitive Economics: Trust, Perception, and the Behavioral Basis of Value

While econophysics and information theory provide quantitative architecture, currency

must also be grounded in human behavior [10], [13], [19]. Cognitive economics contributes a crucial insight:

**Value is not only objective; it is cognitively constructed through shared trust** [14], [15].

Even fiat currencies function because societies collectively believe in their stability. IBC strengthens this cognitive foundation by anchoring value in:

- Verified truth rather than institutional decree.
- Transparent information provenance rather than opaque authority.
- Ethical validation rather than arbitrary consensus.

Three behavioral mechanisms underpin IBC's legitimacy:

### A. Expectation Stabilization

Trusted information reduces cognitive uncertainty. When expectations stabilize, economic fluctuations decrease, enhancing system resilience [11].

### B. Epistemic Trust Networks

Unlike cryptographic trustlessness, IBC fosters **epistemic trust**, where value emerges from credible information intermediaries—research institutions, validators, auditors, and AI-based verification systems.

### C. Cognitive Load Reduction

Information chaos imposes cognitive strain on individuals and institutions. IBC reduces this load by ensuring that monetizable information is already filtered, validated, and contextualized [14], [15]. This aligns with the rational inattention hypothesis, which argues that individuals optimize decisions under limited information-processing capacity [16].

## 3.4 The Convergence of the Three Domains

IBC emerges at the intersection of:

- **Econophysics** → Understanding systemic stability and entropy dynamics
- **Information Theory** → Measuring uncertainty reduction quantitatively
- **Cognitive Economics** → Anchoring value in trust, perception, and behavior

Together, these domains justify a monetary system where:

- Information is not merely symbolic but **structurally meaningful**.
- Currency generation is tied to **epistemic integrity rather than institutional authority or engineered scarcity.**
- Stability emerges from **truth**, not authority or scarcity.

In this sense, IBC is not a reinvention of money but a **reconceptualization of value** suited to an age defined by data, networks, and epistemic fragility.

## 4. Architecture of an Information-Backed Currency System

The operational viability of Information-Backed Currency (IBC) depends on a coherent architecture that transforms raw data into verified, ethically aligned information capable of supporting monetary value [13] with currency issuance governed by the entropy-reduction and verification logic formalized through the ERI–VVU framework introduced in Section 2.4. The architecture consists of **four interlinked layers**:

1. The **Information Ingestion Layer**
2. The **Verification and Validation Layer**
3. The **Ethical Oversight Layer (Dharma Protocol)**
4. The **Monetization and Issuance Layer**

Together, these layers constitute a transparent, resilient, and epistemically grounded economic infrastructure. The design is intentionally interdisciplinary—drawing on cryptographic verification, formal epistemology, digital governance, and macroeconomic modeling—to support the emergence of a next-generation monetary ecosystem.

## 4.1 Layer I: Information Ingestion Layer (Sources → Inputs → Classification)

The IBC system begins with the ingestion of diverse information inputs. Unlike commodity-based systems that rely on static reserves, IBC sources its foundational "value substance" from **verified information events** across sectors.

### 4.1.1 Multi-Source Information Streams

Information enters the system from four primary categories:

- **Scientific and technical knowledge**
  Peer-reviewed research, engineering datasets, validated models, reproducible results.
- **Socio-economic and institutional data**
  Audited financial statements, census data, supply chain records, public governance datasets.
- **Environmental and infrastructure intelligence**
  Satellite data, climatological records, resource inventory assessments.
- **Cultural, civic, and community-generated information**
  Verified participatory data, decentralized knowledge contributions, public deliberation outputs.

### 4.1.2 Raw Data vs. Information Candidates

Not all data qualifies for verification. Data must satisfy minimal criteria to be considered an *information candidate*:

- Source identifiability
- Contextual metadata
- Temporal alignment
- Potential for entropy reduction

This layer acts as a filter, ensuring that only data with informational potential enters the verification pipeline.

## 4.2 Layer II: Verification & Validation Layer (Truth as Technical Function)

This is the **core engine** of IBC: information becomes economically valuable only when it is authenticated, contextualized, and shown to reduce uncertainty.

### 4.2.1 Multi-Path Verification Network (MVN)

Information must pass through **independent verification pathways**, which may include:

- Machine learning–based model triangulation
- Cross-institutional audits
- Citizen verification through consensus-driven platforms
- Expert committee scrutiny
- Cryptographic provenance checks

The independence of pathways is crucial: IBC treats validation as *epistemic redundancy*, ensuring that the truth arises from corroboration, not authority.

### 4.2.2 Entropy Reduction Index (ERI)

ERI quantifies how much new information decreases uncertainty within a particular domain.
Examples:

- A new agricultural yield forecast reduces market volatility
- A validated climate model improves risk pricing
- A peer-reviewed medical study reduces uncertainty around health decisions

ERI functions as the quantitative backbone for monetary issuance.

### 4.2.3 Reproducibility and Contextual Integrity

Information must be:

- **Reproducible** (independent replication)
- **Contextually aligned** (consistent within its domain)
- **Noise-filtered** (high SNR)
- **Bias-screened** (evaluated for distortion)

Only information that passes these tests becomes *verification-grade*.

## 4.3 Layer III: Ethical Oversight — The Dharma Protocol

IBC introduces an unprecedented addition to monetary architecture: **ethical validation** as a precondition for currency backing. This prevents ethically corrosive, harmful, or manipulative information from entering the currency system.

### 4.3.1 Ethical Gatekeeping

The Dharma Protocol evaluates whether information:

- Contributes to social stability
- Avoids harm or systemic manipulation
- Respects privacy, autonomy, and fairness
- Maintains alignment with broad human welfare
- Reduces—not generates—collective vulnerability

### 4.3.2 Normative–Technical Coupling

The Dharma Protocol integrates philosophy with computation:

- Normative reasoning → purpose, intent, societal effect
- Technical evaluation → bias detection, manipulation scores, impact modeling

This ensures ethical intelligence is *embedded*, not externally appended.

### 4.3.3 Preventing Weaponized Information

Under this protocol, harmful information—misinformation, incitement, predatory data—is disqualified from generating economic value.
This reverses a current paradox: today, disinformation is economically profitable; IBC structurally eliminates that incentive.

The Dharma Protocol must not be a static rulebook defined by a single authority. Instead, it must function as a transparent, decentralized governance mechanism where the ethical parameters for currency-grade information are continuously contested, updated, and audited by stakeholders within the ecosystem.

Rather than simply asking "Is this information 'good'?", the protocol asks: "Has this information passed through the established, transparently governed consensus mechanisms for ethical vetting?"

## 4.4 Layer IV: Monetization and Issuance Layer

Once information is verified and ethically validated, it becomes eligible for monetization.

### 4.4.1 Verification Value Unit (VVU)

VVU is the fundamental unit of value in IBC. It represents the **amount of uncertainty reduction** achieved by a verified information event.

Currency issuance is proportional to cumulative VVUs entering the system.

### 4.4.2 Dynamic Issuance Mechanism

Unlike fiat (policy-driven) or crypto (algorithm-driven), IBC issuance is **information-driven**:

- More verified information → controlled expansion of currency supply
- Less verified information → issuance slows automatically
- Manipulated information → issuance halts or reverses (via penalties)

This contributes to macroeconomic stability by tying money supply to *epistemic productivity*.

### 4.4.3 Economic Circulation and Applications

IBC can be used for:

- Public budgeting
- International settlements
- Research funding
- Social credit creation
- Crisis forecasting and resilience modeling

Because IBC is grounded in information transparency, it is resistant to speculative bubbles and corruption.

### 4.4.4 Worked Example: Valuing an Information Event

To illustrate how the theoretical frameworks of information theory and econophysics translate into monetary issuance in practice, we consider a hypothetical scenario involving global food security—a domain characterized by high uncertainty and significant economic consequence [9], [12].

### I. The Pre-Information State (High Entropy)

The global commodities market faces high uncertainty regarding the upcoming season's wheat yield in a critical breadbasket region due to conflicting early weather reports. In information theory terms, the probabilistic distribution of potential outcomes is wide (e.g., predictions ranging from 600 million to 800 million metric tons). This state represents high Shannon entropy (H) within that specific economic subsystem [12].

### II. Data Ingestion and Multi-Path Verification

A massive dataset enters the **Information Ingestion Layer** combining high-resolution satellite imagery (environmental intelligence) with ground-sensor data from regional agricultural institutes (scientific/technical knowledge).

This raw data is processed by the **Verification and Validation Layer** through a Multi-Path Verification Network (MVN):

- **Path A:** Machine learning models analyze satellite data for vegetative health, screening for biases [19].

- **Path B:** Cross-institutional audits confirm the calibration and provenance of the ground sensors.

- **Path C:** The findings are checked for reproducibility against independent climate models.

Only once these independent pathways converge does the data transition from "noise" to a validated "signal" with a high signal-to-noise ratio (SNR) [5], [12].

### III. Calculating the Entropy Reduction Index (ERI)

The system now quantifies the impact of this verified signal. The validated data definitively narrows the probabilistic range of the wheat yield forecast from the previous 600–800MT range to a highly confident 740–760MT range.

This measurable collapse in the range of probable outcomes constitutes the reduction of uncertainty. The difference between the pre-information entropy state and the post-verification entropy state is calculated. This delta is the **Entropy Reduction Index (ERI)** score for this specific information event.

### IV. Determining VVUs and Currency Issuance

The ERI score is mapped to **Verification Value Units (VVUs)**. The VVU represents the standardized economic weight of that specific amount of uncertainty reduction.

Because this information event significantly stabilizes expectations across a major economic sector, it yields a high VVU score. The total quantity of new IBC currency issued into the ecosystem at that moment is directly proportional to the cumulative VVUs generated by this verified event.

## 4.5 Systemic Properties of IBC Architecture

The four-layer structure yields unique systemic advantages:

- **Resilience** — anchored in verified truth, not political volatility
- **Transparency** — traceable information provenance and validation
- **Ethical governance** — harmful information cannot inflate value
- **Scalability** — information grows without resource depletion
- **Decentralized trust** — no single institution controls value formation
- **Lower systemic risk** — reduced information asymmetry reduces economic instability [1].

The IBC architecture therefore constitutes a **novel class of monetary infrastructure**: one that is informational, ethical, transparent, and inherently aligned with planetary-scale digital governance.

## 5. Policy Implications and Futures of an Information-Backed Monetary System

The transition toward an Information-Backed Currency (IBC) represents more than a technological innovation; it constitutes a structural reorientation of how societies conceptualize value, trust, and economic governance. For policymakers and strategic institutions, the rise of information-centric monetary architectures demands a forward-looking regulatory stance that is both adaptive and principled [6]. This section outlines the key policy considerations and future pathways that would determine the feasibility, legitimacy, and global impact of IBC systems.

## 5.1 Regulatory Infrastructure for Information Integrity

An IBC ecosystem presupposes a high-fidelity information environment, making the quality, provenance, and verifiability of data a matter of monetary stability. Policymakers must, therefore, establish:

- **National Information Standards (NIS):** Uniform criteria for data authenticity, metadata structuring, audit trails, and permissible transformations.
- **Independent Information Validation Authorities (IVAs):** Technically capable bodies responsible for certifying information inputs that can qualify as currency-backing assets.
- **Interoperability Protocols:** Cross-sector and cross-border standards ensuring that authenticated information remains usable in multilayered financial and administrative systems.

These foundational elements transform information from a loosely governed digital by-product into a regulated monetary substrate.

## 5.2 Legal Status of Information as an Economic Asset

Recognizing authenticated information as a value-bearing economic asset requires a recalibration of legal frameworks. Key policy adjustments include:

- **Defining Ownership and Economic Rights:** Clarifying whether

information value resides with creators, processors, platforms, or public institutions.
- **Establishing Accountability Chains:** Assigning liability for misinformation, data tampering, or structural risks introduced into IBC circulation.
- **Embedding Information Rights in Monetary Law:** Updating financial statutes to include information-based collateral, reserves, and asset classification norms.

This legal realignment elevates information to the status of a protected and tradable economic resource.

## 5.3 Risk Governance and Systemic Stability

An information-centric monetary system introduces novel risk categories—epistemic risk, algorithmic risk, and provenance corruption. To ensure resilience:

- **Multilayer Risk Scoring:** Central institutions must adopt dynamic risk-scoring models that evaluate the volatility, reliability, and origination complexity of information assets.
- **Redundancy through Distributed Validation:** Avoiding concentration of verification power by embedding distributed consensus models that prevent single-point failures.
- **Adaptive Stress Testing:** Periodic scenario-based tests using synthetic misinformation shocks, network disruptions, or algorithmic anomalies to assess stability thresholds.

Risk governance in IBC is not merely financial; it is informational and epistemological.

## 5.4 Geopolitical and Macroeconomic Considerations

The adoption of IBC has far-reaching geopolitical implications:

- **Digital Sovereignty:** Nations may seek control over domestic information reserves comparable to how they protect critical minerals or energy assets.
- **Competitive Advantage:** Economies with robust information governance infrastructures may emerge as global hubs for IBC-driven financial innovation.
- **Cross-Border Alignment:** IBC necessitates new multilateral frameworks analogous to WTO or FATF—focused on information trade, validation, and ethical compliance.

IBC may, therefore, reshape global influence by elevating informational capital to a strategic national resource.

## 5.5 Ethical and Governance Frameworks for Information Value

IBC intersects directly with questions of autonomy, privacy, and fairness. Because the Dharma Protocol has the power to demonetize information it deems harmful, it introduces a new critical risk: **ideological capture of the monetary system.**

To prevent the ethical layer from becoming a tool for authoritarian control or partisan advantage, policymakers must ensure:

**Immutable Audit Trails for Ethical Decisions:** Any decision by the Dharma Protocol to disqualify information must be cryptographically recorded and publicly accessible, allowing civil society to review the exact rationale used to label information as "harmful" or "manipulative" [8].

- **Pluralistic Ethical Standards:** The system may need to accommodate competing ethical frameworks rather than imposing a single global standard. Different regional or sectoral IBC implementations might operate under different, transparently stated governance charters.
- **Separation of Powers:** The institutions responsible for technical verification (Layer II) must remain operationally distinct from those responsible for ethical oversight (Layer III) to prevent conflicts of interest.

These frameworks are essential for preventing asymmetries of power and safeguarding democratic legitimacy.

## 5.6 Futures Scenarios for IBC Adoption

Several plausible futures shape the trajectory of IBC systems:

1. **Incremental Integration:** Central banks integrate verified information assets into existing reserve structures, using IBC selectively for cross-border settlements.
2. **Dual-Layer Hybrid Currency Systems:** Fiat remains the transactional medium, while IBC governs long-term value stabilization and economic forecasting.
3. **Full-Scale Information-Centric Economies:** Entire monetary ecosystems evolve where information generation, validation, and stewardship become the primary engines of value creation.
4. **Fragmented Adoption:** Competing regional blocs develop divergent IBC standards, creating a multipolar monetary landscape.

Each scenario carries distinct regulatory demands and socio-economic opportunities.

## 5.7 Strategic Imperatives for Policymakers

To responsibly explore IBC, governments and think-tanks should prioritize:

- **Pilot Programs:** Controlled regulatory sandboxes testing limited-scope IBC applications in public finance or supply-chain governance.
- **Institutional Capacity Building:** Developing expertise in data economics, cryptographic validation, and epistemic risk management.
- **Long-Horizon Roadmapping:** Anticipating the shift toward information-centric governance and aligning national strategies accordingly.

The IBC paradigm signals a pivot from industrial-era monetary logic to a future where information is the foremost asset class.

## 6. Conclusion

The emerging paradigm of an Information-Backed Currency (IBC) signals a profound shift in how societies conceptualize value, govern economic systems, and manage collective trust. As the global economy becomes increasingly mediated by artificial intelligence, digital infrastructures, and knowledge-intensive production systems, the foundational assumptions of traditional monetary architectures grow progressively misaligned with the mechanisms that now generate economic and strategic power. IBC offers a conceptual framework that responds to this misalignment by positioning verified, high-quality information as a formal economic asset capable of supporting a resilient, transparent, and adaptive monetary ecosystem.

This paper has articulated the theoretical motivations for IBC, developed its structural components, and analyzed the governance mechanisms necessary to maintain its legitimacy and stability. By reframing information as a measurable, auditable, and value-bearing resource, IBC extends beyond speculative digital currencies or algorithmic monetary tools; it forms the basis for a new class of economic infrastructure grounded in epistemic integrity. In doing so, it addresses a critical deficit in contemporary economic systems—the lack of robust mechanisms to quantify and incorporate information quality into value formation processes.

The policy implications are far-reaching. IBC requires regulatory frameworks that treat information not merely as a digital commodity but as a strategic resource central to national competitiveness and institutional resilience. It demands new legal definitions of ownership, provenance, and liability. It also introduces novel categories of systemic risk that policymakers must anticipate and manage through technocratic governance and global coordination.

At a broader level, IBC challenges think-tanks, academic institutions, and governments to rethink the architecture of trust in a world where information abundance coexists with information fragility. The future of economic stability may not hinge solely on central bank interventions or macroeconomic modeling but on the governance of information flows, the credibility of validation processes, and the ethical frameworks that shape digital ecosystems.

While the model remains conceptual, the trajectory it outlines is already visible across sectors—from knowledge-based industries and precision governance systems to AI-driven markets and data-dependent public institutions. IBC provides a lens through which these developments can be integrated into a unified monetary philosophy capable of guiding long-term policy design.

Ultimately, the value of an Information-Backed Currency lies not only in its potential as a monetary instrument but in its capacity to serve as an intellectual bridge between economics, information science, political theory, and ethics. Its feasibility will depend on interdisciplinary collaboration, rigorous experimentation, and a willingness to reimagine how societies anchor trust in the age of intelligent information.

----------------------------